\documentclass[aps,pra,nobibnotes,twocolumn,superscriptaddress]{revtex4-1}
\usepackage{graphicx}
\usepackage{mathrsfs}
\usepackage{bm}
\usepackage{amsmath}
\usepackage{dcolumn}
\usepackage{epstopdf}
\usepackage{dsfont}
\usepackage{amssymb}
\usepackage{tabularx}
\usepackage{array}
\usepackage{float}
\usepackage{color}
\usepackage{epstopdf}
\usepackage{mathrsfs}
\usepackage[colorlinks, linkcolor=blue,anchorcolor=blue,citecolor=blue,urlcolor=blue]{hyperref}

\begin{document}

\preprint{APS/123-QED}

\title{Magnon bands in twisted bilayer honeycomb quantum magnets}
\author{Xingchuan Zhu}
\affiliation{Department of Physics, Beijing Normal University, Beijing, 100875, China}

\author{Huaiming Guo}
\email{hmguo@buaa.edu.cn}
\affiliation{Department of Physics, Key Laboratory of Micro-Nano Measurement-Manipulation and Physics (Ministry of Education), Beihang University,
Beijing, 100191, China}

\author{Shiping Feng}
\affiliation{Department of Physics, Beijing Normal University, Beijing, 100875, China}


\begin{abstract}

We study the magnon bands of twisted bilayer honeycomb quantum magnets using linear spin wave theory. Although the interlayer coupling can be ferromagnetic or antiferromagnetic, we keep the intralayer one ferromagnetic to avoid possible frustration.
For the interlayer ferromagnetic case, we find the magnon bands have similar features with the corresponding electronic energy spectrums.
Although the linear dispersions near the Dirac points are preserved in the magnon bands of twisted bilayer magnets, their slopes are graduately reduced with the decrease of the twist angles. On the other hand, the interlayer antiferromagnetic couplings generate quite different magnon spectra. The two single-layered magnon spectra are usually undecoupled due to the opposite orientations of the spins in the two layers. We also develop a low-energy continuous theory for very small twist angles, which has been verified to fit well with the exact tight-binding calculations. Our results may be experimentally observed due to the rapid progress in two-dimensional magnetic materials.

\end{abstract}

\maketitle

\section{Introduction}

The recent discovery of the correlated insulator and unconventional superconductivity in twisted bilayer graphene has attracted intense interests\cite{Cao2018,Xie2020,Saito2020,Yankowitz1059,Cao201804,Roy2019,Wu2020,Liu2018}.
By rotating the layers to a small angle, a misalignment induced in the bilayer system introduces a long-period moir\'{e} superlattice. Such a superlattice therefore modulates the electronic structure, and leads to nearly flat band at the magic angles. These bands become so narrow that the electron-electron correlations dominate over kinetic energy, giving rise to the above correlated quantum phases\cite{Balents2020}. Since twisted bilayer graphene is relatively simple and highly tunable, it is anticipated that this system can serve as an ideal platform to investigate the strongly-correlated physics.

Moir\'{e} superlattices have also been employed to engineer flat bands in other two-dimensional (2D) materials, such as twisted double-bilayer graphene, trilayer graphene, and twisted bilayer of the transition metal dichalcogenide, where rich correlation phenomena have been revealed\cite{Chen2019,Li2019,Chen201903,Shi2020,Wang2020,An2020,Shen2020,Burg2019,Chebrolu2019,Culchac2020}. While research has been primarily centered on twisted electronic systems, there appear theoretical studies addressing the remarkable physical properties in twisted Kitaev bilayers\cite{May2020} and twisted optical lattices\cite{Luo2020,Gonz2019}. It is accepting that twisting is a simple and general approach to creating exotic quantum matter.

Recently rapid progress in 2D magnetic atomic crystals has been made\cite{Park2016,burch2018,gibertini2019}. Various monolayer and multilayer van der Waals (vdW) magnetic materials have been discoveried, and different kinds of ferromagnetic(FM) and antiferromagnetic(AFM) orders have been observed\cite{Sharpe605,Bultinck2020,Seo2019,Wu202004,Alavirad2019,Thomson2018,Gu2020,HUANG2019310,Zhang2019}.
The 2D vdW magnetic materials have a huge potential to create novel functional devices, and have important applications in the next-generation nanoelectronics\cite{novoselov2016,Li2019p,Wang2020p,Ningrum2020,Ahn2020}.
Of course, they are also important in fundamental research. The physics can be described by a generalized Heisenberg spin Hamiltonian, and a broad range of parameter regimes can be realized in a rich variety of materials. All three spin Hamiltonians, i.e. Ising, $\textrm{XY}$ and Heisenberg models, can be recovered in specific limits\cite{yuliu2017,yuliu2018,Kim2019,C9CP04685B}. Besides, external perturbations, such as gating, strain et,al., can further tune the range of model parameters\cite{huang2018,yuyan,C8CP07067A,Zheng2014,Zhang2016}. These make the magnetic 2D materials into an ideal platform to examine the well-established theories \cite{onsager1944,mermin1966,Kosterlitz1973} and explore new quantum phases\cite{balents2010,chang2013,lee2018}.

Motivated by these advances in the studies of the twisted 2D materials and 2D magnetic materials, we study how the twist alters the magnon bands in twisted bilayer honeycomb quantum magnets here. We keep the intralayer FM coupling, and study the magnon bands with interlayer FM and AFM couplings in AA(AB)-stacked and twisted bilayer quantum magnets. Our study is closely related to the rapid experimental progress in two-dimensional magnetic materials.

\section{AA-stacked honeycomb magnets}
Twisted bilayers are characterized by non-zero angle $\theta$ between two honeycomb layers. Since the twist may be performed based on AA-stacked bilayer, we start from the spin-$1/2$ Heisenberg model on this geometry, which can be written as,
\begin{eqnarray}\label{eq4}
H=-J\sum_{\langle ij\rangle,\ell}\textbf{S}_{i,\ell}\cdot \textbf{S}_{j,\ell}-J_{\bot}\sum_{\langle ij\rangle}\textbf{S}_{i,2}\cdot \textbf{S}_{j,1},
\end{eqnarray}
where $\textbf{S}_{i,\ell}=(S^{x}_{i,\ell},S^{y}_{i,\ell},S^{z}_{i,\ell})$ is the spin-$1/2$ operator at site $i$ in layer $\ell=1,2$; the summation runs over nearest-neighbor sites $\langle ij\rangle$; $J$ ($J_{\bot}$) is the intralayer (interlayer) coupling constant, and we first consider the FM case, i.e., $J>0$ and $J_{\bot}>0$.

Using Holstein-Primakoff (HP) transformation, the spin operators are expressed in term of bosonic creation and annihilation operators. In the FM case, the transformaiton in the linear spin-wave theory is defined as,
\begin{align}\label{eq5}
S^+_{i}=\sqrt{2S}a_{i},
S^z_{i}=S-a^{\dagger}_{i}a_{i}.
\end{align}
After ignoring a constant and four-operator terms, the resulting bosonic tight binding Hamiltonian becomes,
\begin{eqnarray}\label{eq7}
H_{FM}^{AA}=&-&J\sum_{\langle ij\rangle,\ell} \frac{1}{2}(a^{\dagger}_{i,\ell}a_{j,\ell}+a_{i,\ell}a^{\dagger}_{j,\ell}) \\ \nonumber
&-&J_{\bot}\sum_{i} \frac{1}{2}(a^{\dagger}_{i,1}a_{i,2}+a_{i,1}a^{\dagger}_{i,2})\\ \nonumber
&+&\frac{3J+J_{\bot}}{2}\sum_{i,\ell}a^{\dagger}_{i,\ell}a_{i,\ell}.
\end{eqnarray}
Performing a Fourier transformation, the above Hamiltonian writes as $H=\sum_{\bf k}\psi^{\dagger}_{\bf k}{\cal H}_{FM}({\bf k})\psi_{\bf k}$, where $\psi_{\bf k}=\{ a_{A,1}({\bf k}), a_{B,1}({\bf k}),a_{A,2}({\bf k}), a_{B,2}({\bf k}) \}^{T}$ is the basis, and
\begin{eqnarray}\label{eq8}
{\cal H}_{FM}^{AA}({\bf k})={\cal H}_{0}^{AA}({\bf k})+\frac{3J+J_{\bot}}{2},
\end{eqnarray}
with
\begin{eqnarray}\label{eq8}
{\cal H}_{0}^{AA}({\bf k})=\left(
                    \begin{array}{cccc}
                      0 & f({\bf k}) & -\frac{J_{\bot}}{2} & 0 \\
                      f^*({\bf k}) & 0 & 0 & -\frac{J_{\bot}}{2} \\
                      -\frac{J_{\bot}}{2} & 0 & 0 & f({\bf k}) \\
                      0 & -\frac{J_{\bot}}{2} & f^{*}({\bf k}) & 0 \\
                    \end{array}
                  \right),
\end{eqnarray}
and $f({\bf k})=-\frac{J}{2}(1+e^{-i{\bf k}\cdot{\bf a}_1}+e^{-i{\bf k}\cdot{\bf a}_2})$ [${\bf a}_1=(\sqrt{3},0),{\bf a}_2=(\sqrt{3}/2,3/2)$ the primitive vectors].

With a unitary transformation $\psi_{\bf k}=U({\bf k})\phi_{\bf k}$ ($U({\bf k})^{\dagger}U({\bf k})=\mathbb{I}$),
the above Hamiltonian is directly diagonalized, and the spectrum contains four branches: $\frac{3J+J_{\bot}}{2}\pm \frac{J_{\bot}}{2}\pm |f({\bf k})|$. Compared it to that of single-layer honeycomb magnet, the Dirac points are shifted upward and downward by $\frac{J_{\bot}}{2}$, which is exactly the same as that of AA-stacked graphene except for an overall translation of $\frac{3J+J_{\bot}}{2}$.
\begin{figure}[htbp]
\centering \includegraphics[width=8.8cm]{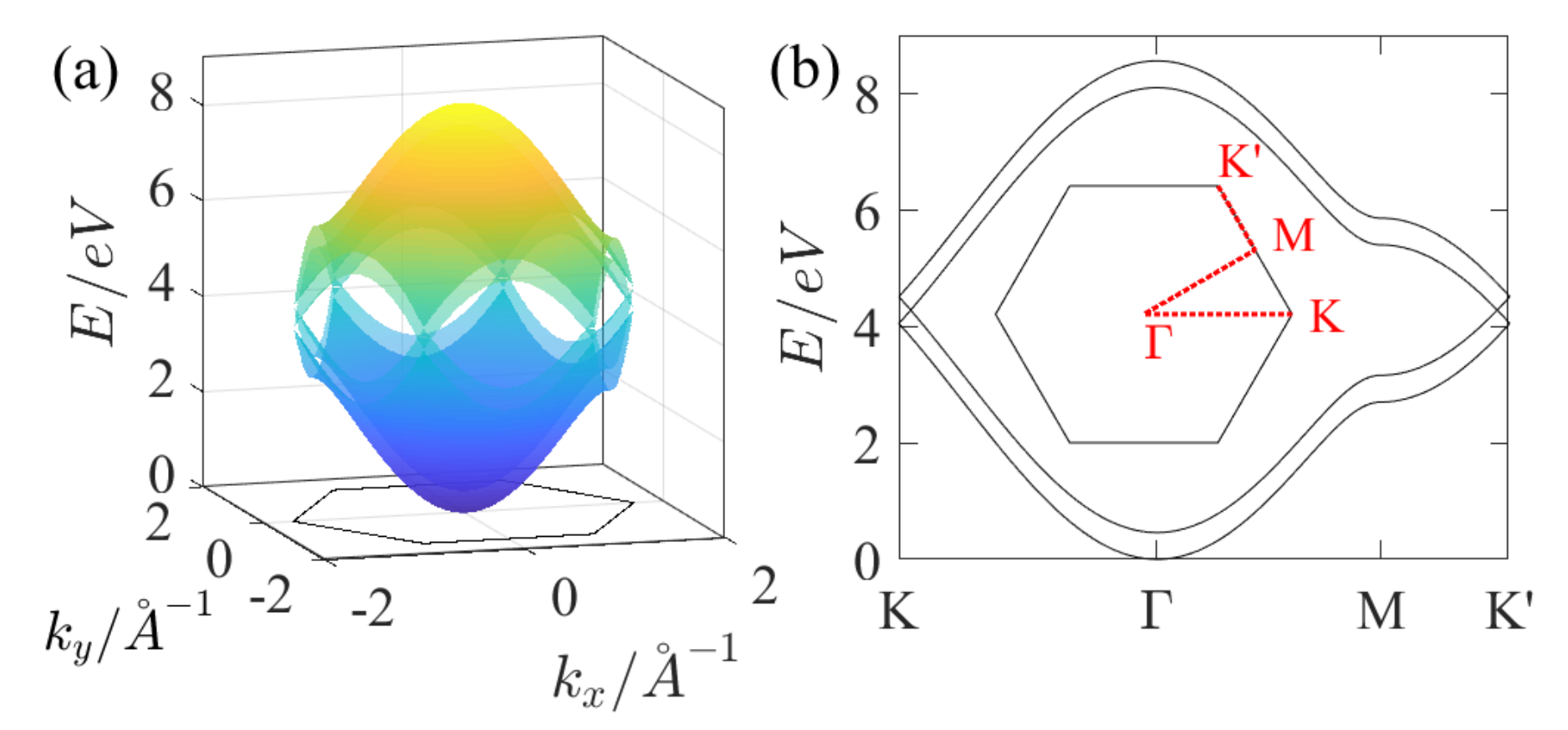} \caption{(a) The magnon band structure of AA-stacked FM bilayer. (b) The magnon bands along the high symmetry points in the first Brillouin zone, which are shown in the inset. The parameters are $J=2.7eV$ and $J_{\bot}=0.46eV$.}
\label{fig1}
\end{figure}

Now we turn to consider the case with interlayer AFM exchange, i.e., $J_{\bot}<0$, while the intralayer coupling remains FM coupling. The following magnetic configuration is assumed: the spins in Layer $1$($2$) are along the positive (negative) $z$-axis. The HP transformation in Layer $1$ is the same as that in Eq.(\ref{eq5}). In Layer $2$, the spin is in the opposite direction, thus the spin operators write as,
\begin{align}\label{eq6}
S^+_{2,i}=a^{\dagger}_{i,2}\sqrt{2S},
S^z_{2,i}=a^{\dagger}_{i,2} a_{i,2}-S.
\end{align}

The resulting bosonic tight binding Hamiltonian becomes,
\begin{eqnarray}\label{eq7}
H_{AFM}^{AA}=&-&J\sum_{\langle ij\rangle,\ell} \frac{1}{2}(a^{\dagger}_{i,\ell}a_{j,\ell}+a_{i,\ell}a^{\dagger}_{j,\ell}) \\ \nonumber
&-&J_{\bot}\sum_{i} \frac{1}{2}(a_{i,1}a_{j,2}+a^{\dagger}_{i,1}a^{\dagger}_{j,2}) \\ \nonumber
&+&\frac{3J-J_{\bot}}{2}\sum_{i,\ell}a^{\dagger}_{i,\ell}a_{i,\ell}.
\end{eqnarray}
Under the basis $\psi_{\bf k}=\{ a_{A,1}({\bf k}), a_{B,1}({\bf k}),a^{\dagger}_{A,2}(-{\bf k}), a^{\dagger}_{B,2}(-{\bf k}) \}^{T}$, the above Hamiltonian in the momentum space can be written as,
\begin{eqnarray}\label{eq8}
{\cal H}_{AFM}^{AA}({\bf k})={\cal H}_{0}^{AA}({\bf k})+\frac{3J-J_{\bot}}{2}, (J_{\bot}<0).
\end{eqnarray}

We use Bogoliubov transformation $\psi_{\bf k}=U({\bf k})\phi_{\bf k}$ to diagonalize the above Hamiltonian $U({\bf k})^{\dagger}{\cal H}_{AFM}^{AA}({\bf k})U({\bf k})=D$, with $D$ is a diagonal matrix. The transformation satisfies $U({\bf k})^{\dagger}s_zU({\bf k})=s_z$ to maintain the commutation relation of bosons. Then we have $s_z{\cal H}_{AFM}^{AA}({\bf k})U({\bf k})=U({\bf k})s_zD$, which means the Bogoliubov transformation $U({\bf k})$ is the eigenvector of $s_z{\cal H}_{AFM}^{AA}$ with the eigenvalue $s_zD$. Thus we obtain the magnon spectrum , $E_{\bf k}^{\pm}=\sqrt{(|f({\bf k})|\pm\frac{3J-J_{\bot}}{2})^2-(\frac{J_{\bot}}{2})^2}$, each of which is two-fold degenerate. For small $J_{\bot}$ and near the Dirac points, {$E_{\bf k}^{\pm}\sim \pm|f({\bf k})|+\frac{3J-J_{\bot}}{2}- \frac{J^2_{\bot}}{4(3J-J_{\bot})}$, and the Dirac point is slightly shifted downward, and the dispersion remains linear. As shown in Fig. \ref{fig2}, the magnon spectrum is almost the same as that of the single-layer FM case, suggesting the magnon excitations from different spin alignments are slightly coupled.

\begin{figure}[htbp]
\centering \includegraphics[width=8.8cm]{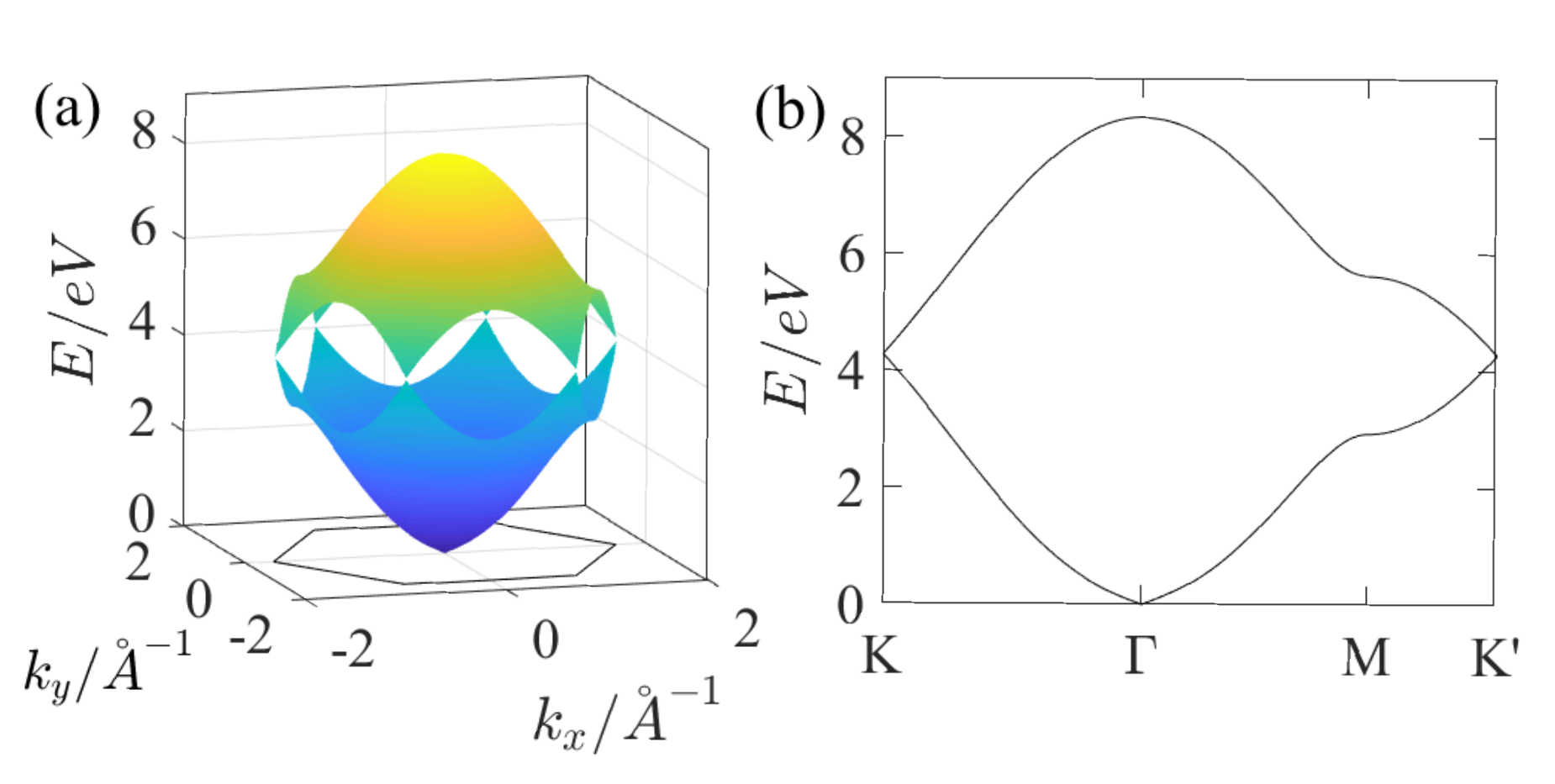} \caption{(a) The magnon band structure of AA-stacked FM bilayer with interlayer AFM exchange. (b) The magnon bands along the high symmetry points in the first Brillouin zone. The parameters are the same as those of Fig. \ref{fig1}.}
\label{fig2}
\end{figure}

\section{AB-STACKED HONEYCOMB MAGNETS}

Next we consider AB-stacked bilayer quantum magnets, which can be viewed as rotating AA-stacked bilayer by $\theta=60^\circ$. In the intralayer and interlayer FM case, the bosonic tight binding Hamiltonian is
\begin{eqnarray}\label{eqABcengjiantieci}
H_{FM}^{AB}=&-&J_{\bot}\sum_{i} \frac{1}{2}(a^{\dagger}_{i_{B},1}a_{i_{A},2}+a_{i_{B},1}a^{\dagger}_{i_{A},2})\\ \nonumber
&+&\frac{J_{\bot}}{2}\sum_{i}(a^{\dagger}_{i_{B},1}a_{i_{B},1}+a^{\dagger}_{i_{A},2}a_{i_{A},2}) \\ \nonumber
&-&J\sum_{\langle ij\rangle,\ell} \frac{1}{2}(a^{\dagger}_{i,\ell}a_{j,\ell}+a_{i,\ell}a^{\dagger}_{j,\ell})+\frac{3J}{2}\sum_{i,\ell}a^{\dagger}_{i,\ell}a_{i,\ell}.
\end{eqnarray}
In the momentum space, it writes as,
\begin{eqnarray}\label{eqABcengjiantieci2}
{\cal H}_{FM}^{AB}({\bf k})={\cal H}_{0}^{AB}({\bf k})+{\cal D}_{FM},
\end{eqnarray}
with
\begin{eqnarray*}\label{eqABcengjiantieci2A}
{\cal H}_{0}^{AB}({\bf k})=\left(
                    \begin{array}{cccc}
                      0 & f({\bf k}) & 0 & 0 \\
                      f^*({\bf k}) & 0 & -\frac{J_{\bot}}{2} & 0 \\
                      0 & -\frac{J_{\bot}}{2} & 0 & f({\bf k}) \\
                      0 & 0 & f^{*}({\bf k}) & 0 \\
                    \end{array}
                  \right),
\end{eqnarray*}
and
\begin{eqnarray*}\label{eqABcengjiantieci2B}
{\cal D}_{FM}=\left(
                    \begin{array}{cccc}
                      \frac{3J}{2} & 0 & 0 & 0 \\
                      0 & \frac{3J+J_{\bot}}{2} & 0 & 0 \\
                      0 & 0 & \frac{3J+J_{\bot}}{2} & 0 \\
                      0 & 0 & 0 & \frac{3J}{2} \\
                    \end{array}
                  \right).
\end{eqnarray*}
The spectrum contains four branches: $E^{\pm}_1=\frac{3J}{2}\pm|f({\bf k})|$ and $E^{\pm}_{2}=\frac{3J+J_{\bot}}{2}\pm \frac{\sqrt{J_{\bot}^2+4|f({\bf k})|^2}}{2}$. $|f({\bf k})|$ is linear in $k$ near the Dirac points $\textrm{K, K'}$, and thus $E^{\pm}_{1}$ cross at $\frac{3J}{2}$ linearly[see Fig.\ref{fig3} (a)]. Meanwhile, $E_2^{+}\sim \frac{3J}{2}+J_{\bot}+\frac{|f({\bf k})|^2}{J_{\bot}}$ and $E_2^{-}\sim \frac{3J}{2}-\frac{|f({\bf k})|^2}{J_{\bot}}$, which are quadratic in $k$ near the Dirac points.

\begin{figure}[htbp]
\centering \includegraphics[width=8.8cm]{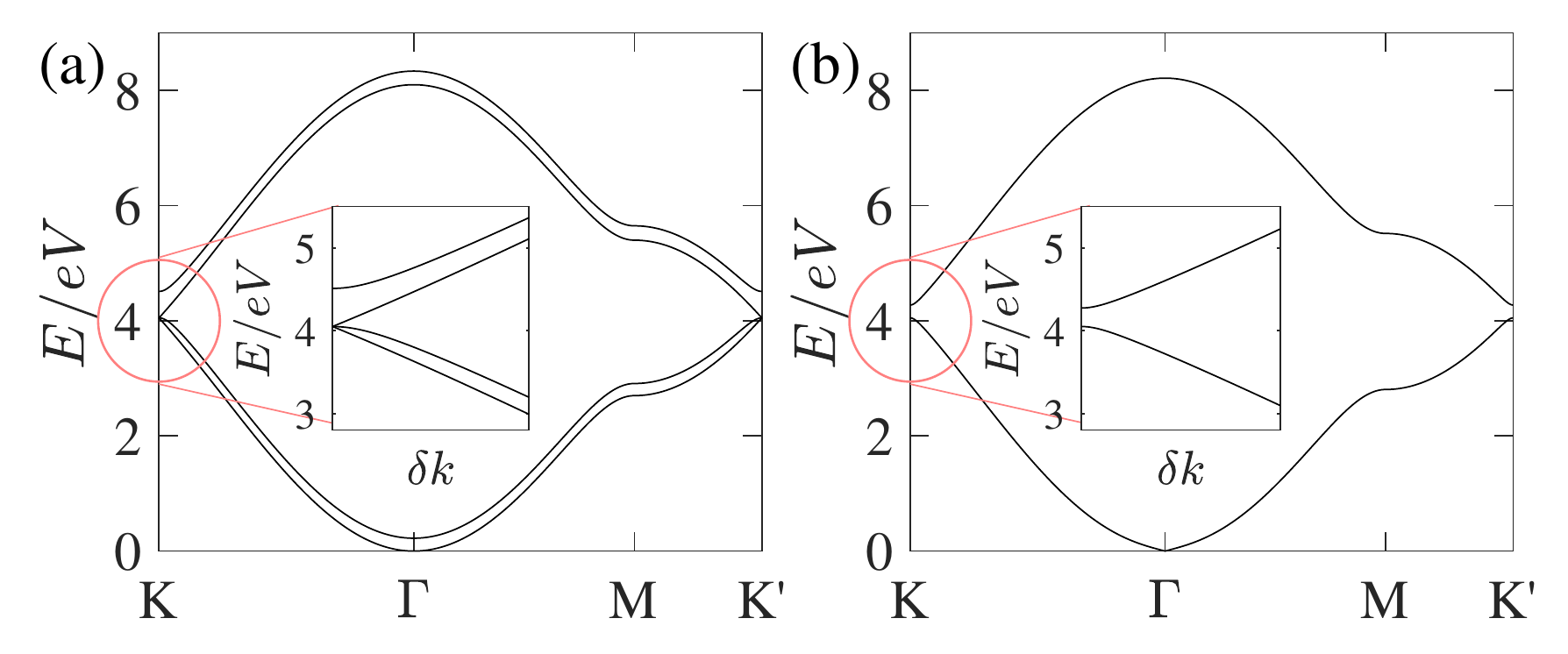} \caption{The magnon bands along the high symmetric lines in the first Brillouin zone for AB-stacked bilayers with: (a) the interlayer FM coupling. (b) the interlayer AFM coupling. Insets of both figures show the enlarged plots of the magnon bands near the Dirac point.}
\label{fig3}
\end{figure}

Then we change interlayer coupling AFM. The bosonic tight binding Hamiltonian becomes,
\begin{eqnarray}\label{eqABcengjianfantieci}
H_{AFM}^{AB}=&-&J_{\bot}\sum_{i} \frac{1}{2}(a^{\dagger}_{i_A,2}a^{\dagger}_{i_B,1}+a_{i_A,2}a_{i_B,1})\\ \nonumber
&-&\frac{J_{\bot}}{2}\sum_{i}(a^{\dagger}_{i_B,1}a_{i_B,1}+a^{\dagger}_{i_A,2}a_{i_A,2}) \\ \nonumber
&-&J\sum_{\langle ij\rangle,\ell} \frac{1}{2}(a^{\dagger}_{i,\ell}a_{j,\ell}+a_{i,\ell}a^{\dagger}_{j,\ell})+\frac{3J}{2}\sum_{i,\ell}a^{\dagger}_{i,\ell}a_{i,\ell}.
\end{eqnarray}
In the momentum space, we have,
\begin{eqnarray}\label{eqABcengjiantieci2}
{\cal H}_{AFM}^{AB}({\bf k})={\cal H}_{0}^{AB}({\bf k})+{\cal D}_{AFM},
\end{eqnarray}
with
\begin{eqnarray*}\label{eqABcengjiantieci2A}
{\cal D}_{AFM}=\left(
                    \begin{array}{cccc}
                      \frac{3J}{2} & 0 & 0 & 0 \\
                      0 & \frac{3J-J_{\bot}}{2} & 0 & 0 \\
                      0 & 0 & \frac{3J-J_{\bot}}{2} & 0 \\
                      0 & 0 & 0 & \frac{3J}{2} \\
                    \end{array}
                  \right),
\end{eqnarray*}
By diagonalizing $s_z{\cal H}_{AFM}^{AB}({\bf k})$, we obtain the magnon spectrum, $E_{\bf k}^{\pm}=\sqrt{\left(\frac{3J}{2}\right)^2 \pm \frac{g({\bf k})}{2}+|f({\bf k})|^2-\left(\frac{3J}{2}\right)\frac{J_{\bot}}{2}}$, with $g({\bf k})=\sqrt{\frac{3J}{2}\left[\left(\frac{3J}{2}\right)J_{\bot}^2-8J_{\bot}|f({\bf k})|^2+16\left(\frac{3J}{2}\right)|f({\bf k})|^2\right]}$. Near the Dirac points, $E^{+}\sim \frac{3J}{2}+\frac{[4(3J/2)-J_{\bot}]|f({\bf k})|^2}{2(3J/2)J_{\bot}}$, and $E^{-}\sim \sqrt{\left(\frac{3J}{2}\right)^2-\left(\frac{3J}{2}\right)J_{\bot}}+\frac{[-4(3J/2)+3J_{\bot}]|f({\bf k})|^2}{2J_{\bot}\sqrt{(3J/2)^2-(3J/2)J_{\bot}}}$, which are quadratic in $k$. The spectrum is gapped on the Dirac points, with the gap size $\frac{3J}{2}-\sqrt{\left(\frac{3J}{2}\right)^2-\left(\frac{3J}{2}\right)J_{\bot}}$.

\begin{figure}[htbp]
\centering \includegraphics[width=8.8cm]{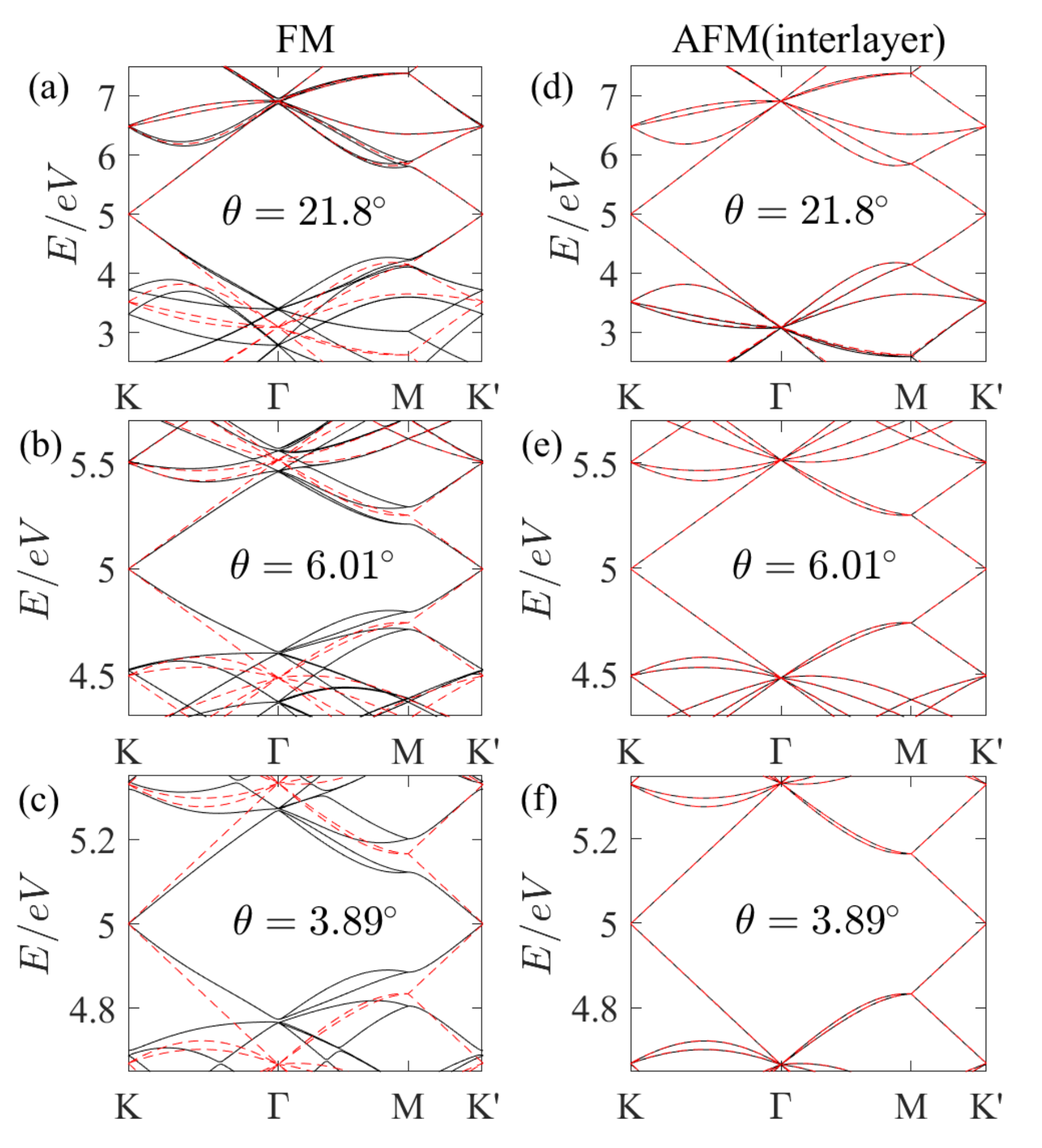} \caption{Magnon band structures of twisted bilayers with twist angles of $21.8^\circ$, $6.01^\circ$ and $3.89^\circ$. (a), (b), (c): the interlayer FM coupling; (d), (e), (f): the interlayer AFM coupling. The intralayer couplings are FM case in all figures. We also plot the magnon spectrums without interlayer coupling (dashed red lines). To compare the Dirac velocity, the Dirac points are moved to the same positions as those in twisted bilayers.}
\label{fig4}
\end{figure}

\section{Twisted bilayer quantum magnets}
By taking any vertical bond in AA-stacked bilayer as the axis, the twisted bilayers are generated by rotating one layer around the axis, while the other layer remains fixed. The periodicity is maintained for commensurate $\theta$, when a certain site of one layer ends up exactly over a site of the other after the rotation. The Bravais lattice of the superstructure is hexagonal with the two unit vectors: $L_{1}=m\vec{a}_{1}+n\vec{a}_2$ and $L_{2}=-n\vec{a}_{1}+(m+n)\vec{a}_2$. Each supercell contains $N=4(m^2+mn+n^2)$ sites\cite{nori2016,helin2017}. The twist angle writes as,
\begin{eqnarray*}
\theta=2 \sin ^{-1} \frac{m-n}{2 \sqrt{m^{2}+n^{2}+m n}}.
\end{eqnarray*}
In the following, we consider the commensurate twist angles $21.8^\circ$, $6.01^\circ$ and $3.89^\circ$, which correspond to $(m,n)=(2,1),(6,5),(9,8)$, respectively. The case of smaller angles have too large supercells to deal with using our computer, and thus instead we calculate the magnon spectrum using the low-energy continuous theory near the Dirac points\cite{Bistritzer2011,castro2012}.

The nearest-neighbor intralayer FM coupling are considered with the strength as the energy scale. The strength of interlayer coupling between sites $r_{i}$ and $r_{j}$ is given by $
J_{i j}=J_{\perp} e^{-\left[\left(\left|r_{i}-r_{i}\right|-d_{0}\right) / \xi\right]}$, where the parameters are set as $J_{\perp}=-0.46$ eV, $d_{0}=$ $0.335 \mathrm{nm},$ and $\xi=0.0453 \mathrm{nm}$.

The magnon spectrums are show in  Fig.\ref{fig4}. Since the $z$-component of each Heisenberg term generates a positive potential in the HP transformation, the bands are shifted upward compared to the cases without the interlayer coupling. The Dirac cones are preserved for the studied angles. In the interlayer FM cases, while the Fermi velocities are almost unchanged for large twist angles ($\theta=21.8^\circ, 6.01^\circ$), there is a clear reduction for small angle $\theta=3.89^\circ$. The behavior is very similar to that of twisted bilayer graphene\cite{morell2010,trambly2010}. It is expected that the magnon bands becomes flatting as the angle is further decreased. In contrast, the spectrum of the interlay AFM case differs little from that of the single-layer quantum FM case, further verifying the spin excitations of opposite spin orientations do not couple with each other.

Now we begin to study the mangon spectrums of small twist angles using the low-energy continuous theory near the valley $\mathbf{K}_{\pm}=\frac{4\pi}{3\sqrt{3}a}(\pm 1,0)$, where $a$ is the lattice constant. Ignoring the overall potential from the z-component of the Heisenberg couplings, the effective bosonic Hamiltonian consists of three terms\cite{Bistritzer2011,castro2012,Koshino2015,Koshino2018,ashvin2019,jhgao2019,liuzhao2019,Balents2019},
\begin{align}\label{eq13}
H_{eff}^{(\pm)}&=\sum_{\mathbf{k}}[ \psi_b^\dagger(\mathbf{k})h^{(\pm)}_{\theta/2}\left(\mathbf{k}-\mathbf{K}^b_{\pm}\right)\psi_b(\mathbf{k}) \\ \nonumber
&+\psi_t^\dagger(\mathbf{k})h^{(\pm)}_{-\theta/2}\left(\mathbf{k}-\mathbf{K}^t_{\pm}\right)\psi_t(\mathbf{k}) \\ \nonumber
&+\sum_{j=1}^3 \left(\psi_b^{\dagger}(\mathbf{k})T^{(\pm)}_j\psi_t(\mathbf{k}+\mathbf{X}^{(\pm)}_j)+h.c.\right)]
\end{align}
with $\mathbf{K}^t_{\pm}=\mathcal{R}_{\theta/2}\mathbf{K}_{\pm}$ and $\mathbf{K}^b_{\pm}=\mathcal{R}_{-\theta/2}\mathbf{K}_{\pm}$, where $\mathcal{R}_{\theta}$ rotates
a vector by angle $\theta$ counterclockwise. $\mathbf{k}$ represents the momenta near the Dirac point $\mathbf{K}_{\pm}$, which writes as $\mathbf{k}_0+m\mathbf{G}_1+n\mathbf{G}_2+\mathbf{Y}_{\pm}$ with $\mathbf{k}_0$ in the moir$\acute{e}$ first Brillouin zone and $\mathbf{Y}_{\pm}=(\mathbf{K}^t_{\pm}+\mathbf{K}^b_{\pm})/2 \pm \mathbf{G}_1/2$. Here, $\mathbf{G}_1=K_{\theta}(1,0)$ and $\mathbf{G}_2=K_{\theta}(-\frac{1}{2},\frac{\sqrt{3}}{2})$ with $K_{\theta}=\frac{8\pi sin(\theta/2)}{3a}$ is the length of the basis vectors in the moir$\acute{e}$ Brillouin zone. In practice, the integers $m, n$ are cut off to finite values, i.e., $m,n=-l,...,l$, and the magnon spectrums converge quickly as the cutoff value $l$ increases. The first two terms describe the isolated FM sheets of the top and bottom layers under the bases
\begin{align}\label{eq14}
\psi_b(\mathbf{k})=
\left[
\begin{array}{c}
\varphi_{A,b}(\mathbf{k})\\
\varphi_{B,b}(\mathbf{k})
\end{array}
\right]
,
\psi_t(\mathbf{k})=
\left[
\begin{array}{c}
\varphi_{A,t}(\mathbf{k})\\
\varphi_{B,t}(\mathbf{k})
\end{array}
\right].
\end{align}
The low-energy Hamiltonian for a layer rotated by an angle $\theta$ is $h^{(\pm)}_{\theta}(\mathbf{k})=\hbar v_f (\mathcal{R}_\theta \mathbf{k})\cdot(\pm \sigma_x,\sigma_y)$, which is a Dirac one. The last term in Eq.(\ref{eq13}) describes hoppings between layers with the hopping matrix
\begin{align}\label{eq15}
T^{(\pm)}_j=\omega
\left[
\begin{array}{cc}
1&e^{\mp\frac{2\pi i}{3}(j-1)}\\
e^{\pm\frac{2\pi i}{3}(j-1)}&1
\end{array}
\right],
\end{align}
and $\mathbf{X}^{(\pm)}_1=(0,0)$, $\mathbf{X}^{(\pm)}_2=\pm(\mathbf{G}_1+\mathbf{G}_2)$ and $\mathbf{X}^{(\pm)}_3=\pm\mathbf{G}_2$. The magnon spectrum is calculated directly, and the result for the $\theta=3.89^{\circ}$ case is shown in Fig. \ref{fig5}. We find the interlayer coupling strength $\omega=0.0439eV$ gives the best match between the continuous theory and exact tight-binding calculation. It is expected that $\omega$ will change with the twisted angle $\theta$. However the determined strength at relatively large $\theta$ provides a good approximation for the cases of smaller angles, which are very hard to access in the tight-binding method.

\begin{figure}[htbp]
\centering \includegraphics[width=6.8cm]{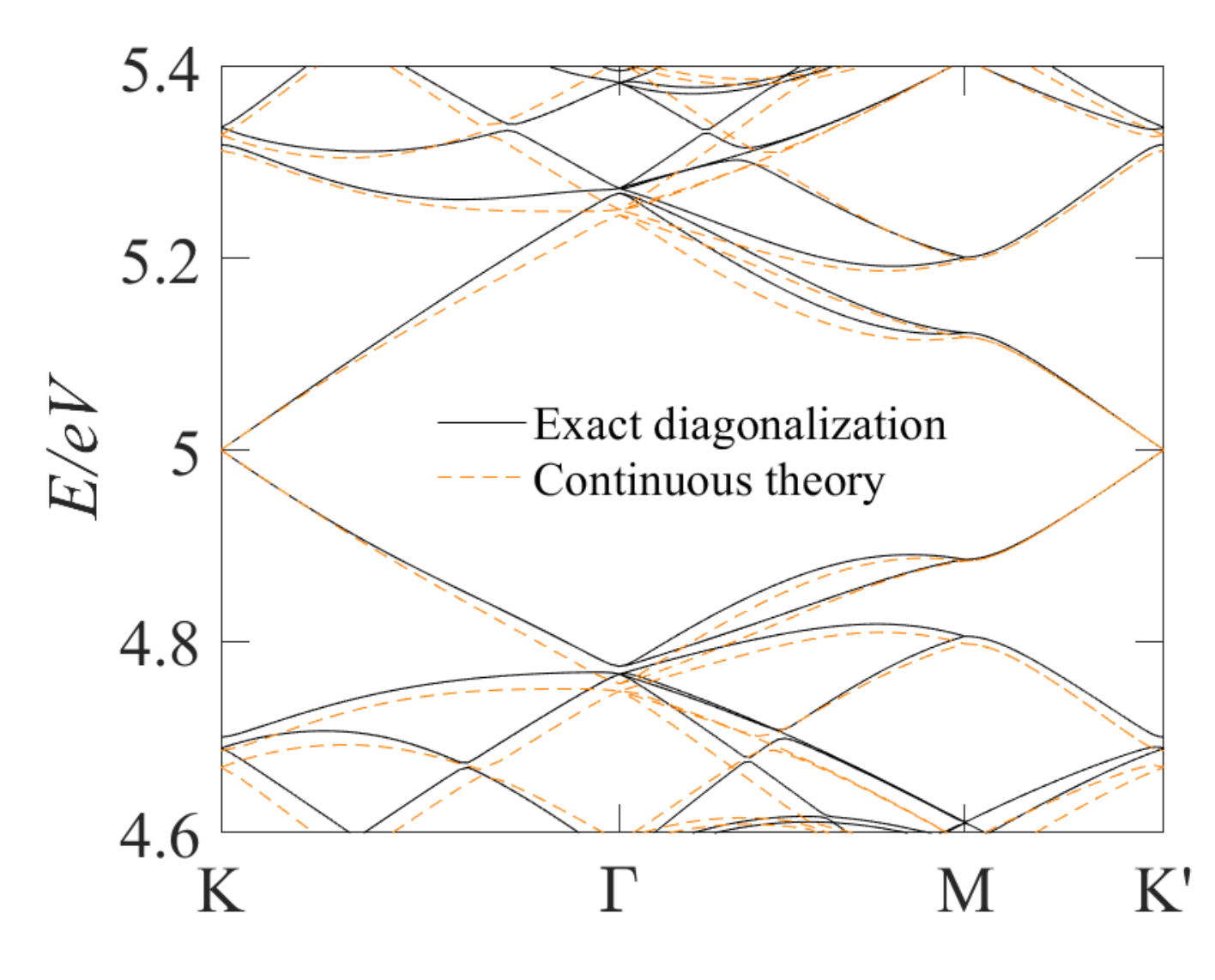} \caption{Comparison of the low-energy magnon bands between the continuous theory and exact tight-binding calculation with the FM interlayer coupling. Here the twist angle is $\theta=3.89^{\circ}$.}
\label{fig5}
\end{figure}

\section{Conclusions}

We study the magnon bands of twisted bilayer honeycomb quantum magnets using linear spin wave theory. Although the interlayer coupling can be FM or AFM, the intralayer one is always kept FM to avoid possible frustration.
For the interlayer FM case, the magnon bands have similar features with the corresponding electronic energy spectrums.
In the AA-stacked magnets, it consists of two single-layered magnon spectra shifted relative to each other by the energy $J_{\bot}$.
For the AB-stacked case, two magnon bands remain linear near the touching point $3J/2$, while the other two becomes parabolic and gapped.
We then show the magnon bands of twisted bilayer magnets, and find that although the linear dispersions near the Dirac points are preserved, their slopes are graduately reduced with the decrease of the twist angles. Nevertheless, the interlayer AFM couplings generate quite different magnon spectra, which may be due to the opposite orientations of the spins in the two layers. The two single-layered magnon spectra are undecoupled in the AA-stacked and the twisted magnets, and the magnon bands are just similar to those of the single-layered magnet. For the AB-stacked magnets, the linear dispersions change to be parabolic and gapped. Finally, we develop a  low-energy continuous theory for very small twist angles, which fits well with the exact tight-binding calculations.

Experimentally there have revealed intrinsic magnetism in various atomically thin crystals. Among them, $CrX_{3} (X=Cl, Br, I)$ is a family of 2D honeycomb quantum magnets. Specifically, intralayer ferromagnetism and interlayer antiferromagnetism has been observed in bilayer $CrI_{3}$\cite{huang2017,Song2018,Klein2018,Thiel2019}. Besides, it has been proposed that the interlayer exchange coupling can be tuned between AFM and FM by changing the interlayer stacking order\cite{dixiao2018}. Thus based on these bilayer honeycomb quantum magnets, our results may be experimentally observed using the state of the art measurements, such as inelastic neutron scattering and magneto-Raman spectroscopy\cite{xdxu2020}.

\section{Acknowledgments}
H.G. acknowledges support from the NSFC grant No.~11774019, the Fundamental Research
Funds for the Central Universities and the HPC resources
at Beihang University.
X.Z. and S.F. are supported by the National Key Research and Development Program of China under Grant No. 2016YFA0300304, and NSFC under Grant Nos. 11974051 and 11734002.

\nocite{*}
\bibliography{magnon_ref}

\appendix

\end{document}